\documentclass[12pt]{iopart}
\usepackage{graphicx}
\usepackage{epsfig}
\usepackage{soul}
\usepackage{iopams}

\begin{document}

\title{Quantum discord for a two-parameter class of states in $2 \otimes d$ quantum systems}

\author{ Mazhar Ali}
\address{ Department of Electrical Engineering, COMSATS Institute of Information Technology, Abbottabad 22060, Pakistan}
\ead{mazharaliawan@yahoo.com}

\begin{abstract}

Quantum discord witnesses the nonclassicality of quantum states even when there is no entanglement in these quantum states. This type of quantum correlation also has some interesting and significant applications in quantum information processing. Quantum discord has been evaluated explicitly only for certain class of two-qubit states. We extend the previous studies to $2 \otimes d$ quantum systems and derive an analytical expression for quantum discord for a two-parameter class of states for $d \geq 3$. We compare quantum discord, classical correlation, and entanglement for qubit-qutrit systems to demonstrate that different measures of quantum correlation are not identical and conceptually different.

\end{abstract}

\submitto{\JPA}

\maketitle

\section{Introduction}

Bipartite quantum states can be divided into separable states and entangled states \cite{Alber-QI2001, Horodecki-RMP-2009}. Entangled states have been shown as a resource for certain tasks in quantum information \cite{NC-QIQC-2000}. Entangled states are nonclassical because they can not be prepared using local operations and classical communication (LOCC). Recently, it was found that there are some quantum correlations namely quantum discord different than entanglement which are also nonclassical and offer some advantage over classical states, for example, quantum non-locality without entanglement \cite{Bennett-PRA59-1999, Horodecki-PRA71-2005, Niset-PRA74-2006}, and broadcasting of quantum states \cite{Piani-PRL100-2008, Piani-PRL102-2009}. In addition, it was shown theoretically \cite{Braunstein-PRL83-1999, Meyer-PRL85-2000, Datta-PRL100-2008}, and later experimentally \cite{Lanyon-PRL101-2008}, that quantum discord is useful for quantum computation \cite{Cui-JPA43-2010}. Therefore, it is desirable to investigate, characterize, and quantify quantum discord and other correlations to have a unified view of quantum and classical correlations \cite{Modi-PRL104-2010}.

Quantum discord is defined only for bipartite systems as this concept relies on definition of quantum mutual information which is defined only for bipartite systems. Quantum mutual information captures the total amount of correlations, both classical and quantum correlation in a given quantum state. It is an information-theoretic measure of the total correlation in a bipartite quantum state \cite{Groisman-PRA72-2005}. In particular, if $\rho^{AB}$ denotes the density operator of a composite bipartite system $AB$, and $\rho^A$ ($\rho^B$) the density operator of part $A$ ($B$), respectively, then the quantum mutual information is defined as
\begin{eqnarray}
\mathcal{I} (\rho^{AB}) = S (\rho^A) + S (\rho^B) - S(\rho^{AB})\, , \label{Eq:QMI}
\end{eqnarray}
where $S(\rho) = - \mathrm{tr} \, ( \rho \, \log_2 \rho )$ is the von Neumann entropy. Moreover, it was shown that quantum mutual information is the maximum amount of information that A(lice) can send securely to B(ob) if a composite correlated quantum state is used as the key for a {\it one-time pad cryptographic system} \cite{Schumacher-PRA74-2006}. Recently it was suggested that correlations in a given quantum state can be split into two parts, that is as a sum of classical correlation $\mathcal{C}(\rho^{AB})$ and quantum correlation $\mathcal{Q} (\rho^{AB})$, that is, $\mathcal{I} (\rho^{AB}) = \mathcal{C} (\rho^{AB}) + \mathcal{Q} (\rho^{AB})$ \cite{Ollivier-PRL88-2001, Vedral-et-al, Luo-PRA77-2008}. The quantum part $\mathcal{Q}$ has been called quantum discord \cite{Ollivier-PRL88-2001}. Quantum discord is not identical to entanglement because separable mixed states (that is, with no entanglement) can have non-zero quantum discord.
 
Quantum discord reflects the nonclassicality of quantum states. Despite its practical applications and importance in understanding the fundamental questions in quantum physics, quantum discord has only been evaluated for the simplest case of qubit-qubit systems and even for restricted but larger class of quantum states \cite{Mali-PRA81-2010}. The main reason is the difficulty of complicated extremization procedure which becomes intractable with growing number of parameters both in von Neumann measurements and the parameters involved with quantum states. For pure states and, surprisingly, for a mixture of Bell states, quantum correlation is exactly equal to entanglement whereas classical correlation attains its maximum value $1$. However, for general two-qubit mixed states, the situation is more complicated. Qubit-qubit entanglement has been characterized and quantified completely whereas quantum discord only for particular cases \cite{Ollivier-PRL88-2001, Vedral-et-al, Luo-PRA77-2008, Mali-PRA81-2010, Li-Luo-PRA-2008, Oppenheim-PRL89-2002, Kaszlikowski-PRL101-2008, Dillenschneider-PRB78-2008, Sarandy-PRA80-2009, Werlang-PRA80-2009}. We slightly extend and generalize some of the previous studies to analytically compute the classical correlation and quantum discord for a two-parameter class of states in $2 \otimes d$ quantum systems with $d \geq 3$. This class of states can be obtained from an arbitrary state by means of LOCC and these states are invariant under all bilateral operations on a $2 \otimes d$ quantum system. We show that classical correlation and quantum discord can be calculated straight forwardly due to the fact that the eigenvalues of the measurement ensemble do not depend on the parameters of the von Neumann measurements. This fact is due to the reason that these states are highly symmetric and this symmetry brings sufficient simplicity in handling the corresponding maximization/minimization procedure. As an example, we study the qubit-qutrit system and compare the classical correlation, quantum discord, and entanglement for various initial states.

This paper is organized as follows. In Section \ref{discord} we discuss quantum discord. We describe the two-parameter class of states in Section \ref{tp-states} and calculate the classical correlation and quantum discord for them. In Section \ref{relation}, we apply the results for various initial states of qubit-qutrit system and study the relation between the classical correlation, quantum discord, and entanglement. We conclude our work in Section \ref{conclusion}. An Appendix present the details of transforming an arbitrary state in $2 \otimes d$ quantum systems by local operation and classical communication (LOCC) to two-parameter class of quantum states discussed in this paper.

\section{Quantum discord}\label{discord}

In order to quantify quantum discord, Ollivier and Zurek \cite{Ollivier-PRL88-2001} suggested the use of von Neumann type measurements which consist of one-dimensional projectors that sum to the identity operator. Let the projection operators $\{ B_k\}$ describe a von Neumann measurement for subsystem $B$ only, then the conditional density operator $\rho_k$ associated with the measurement result $k$ is 
\begin{eqnarray}
\rho_k = \frac{1}{p_k} (I \otimes B_k) \rho (I \otimes B_k) \,,
\end{eqnarray}
where the probability $p_k$ equals $\mathrm{tr} [(I \otimes B_k) \rho (I \otimes B_k)]$.
The quantum conditional entropy with respect to this measurement is given by \cite{Luo-PRA77-2008} 
\begin{eqnarray}
S (\rho | \{B_k\}) := \sum_k p_k \, S(\rho_k) \, ,
\label{Eq:QCE}
\end{eqnarray}
and the associated
quantum mutual information of this measurement is defined as
\begin{eqnarray}
\mathcal{I} (\rho|\{B_k\}) := S (\rho^A) - S(\rho|\{B_k\}) \, . \label{Eq:QMIM} 
\end{eqnarray}
A measure of the resulting classical correlations is provided \cite{Ollivier-PRL88-2001, Vedral-et-al, Luo-PRA77-2008, Mali-PRA81-2010, Li-Luo-PRA-2008} by 
\begin{eqnarray}
\mathcal{C}_B(\rho) := \sup_{\{B_k\}} \, \mathcal{I} (\rho|\{B_k\}) \, . \label{Eq:CC} 
\end{eqnarray}

The real obstacle to compute quantum discord lies in this complicated maximization procedure for calculating the classical correlation because the maximization is to be done over all possible von Neumann measurements of $B$. Now, the quantities $\mathcal{I}(\rho)$ and $\mathcal{C}_B$ may differ, and the difference 
\begin{eqnarray}
\mathcal{Q}_B(\rho) := \mathcal{I}(\rho) - \mathcal{C}_B(\rho) \, ,
\end{eqnarray}
is called quantum discord.

The above definition of quantum discord is not symmetric with respect to parties $A$ and $B$ \cite{Fanchini-PRA81-2010}. Nevertheless, one can swap the role of $A$ and $B$ and get another expression \cite{Bylicka-PRA81-2010} for quantum discord as
\begin{eqnarray}
\mathcal{Q}_A(\rho) := \mathcal{I}(\rho) - \mathcal{C}_A(\rho) \, ,
\end{eqnarray}
where 
\begin{eqnarray}
\mathcal{C}_A(\rho) := \sup_{\{A_k\}} \, \mathcal{I} (\rho|\{A_k\}) \, , \label{Eq:CCa} 
\end{eqnarray}
and $\{ \, A_k \, \}$ describe the von Neumann measurements for subsystem $A$ only. For a given mixed state $\mathcal{Q}_A (\rho) \neq \mathcal{Q}_B(\rho)$, however it is known that $\mathcal{Q}_A (\rho) \, , \, \, \mathcal{Q}_B(\rho) \geq 0$. For pure states, quantum discord coincides with the von Neumann entropy of entanglement. For two-qubit quantum states, the quantification of quantum discord is only available for certain states \cite{Luo-PRA77-2008, Mali-PRA81-2010, Li-Luo-PRA-2008, Oppenheim-PRL89-2002, Kaszlikowski-PRL101-2008, Dillenschneider-PRB78-2008, Sarandy-PRA80-2009, Werlang-PRA80-2009}. It is known that if $\mathcal{Q}_A (\rho) = \mathcal{Q}_B(\rho) = 0$, then $\rho$ is diagonal in the product basis $|i\rangle \otimes |j\rangle$, and 
\begin{eqnarray}
 \rho = \sum_{i,j} \lambda_{ij} \, |i\rangle\langle i| \otimes |j\rangle\langle j| \, ,
\end{eqnarray}
can be represented by the classical joint probability distribution $\lambda_{ij}$. Recently, it was shown \cite{Ferraro-PRA81-2010} that if $\mathcal{Q}_A (\rho) = 0 $, then
\begin{eqnarray}
 [ \, \rho, \, \rho_A \otimes I_B \, ] = 0.
\end{eqnarray}
This necessary condition ensures that if $\rho$ does not commute with $\rho_A \otimes I_B$, then its quantum discord is strictly positive, and $\rho$ is nonclassically correlated. However, the reverse is not true, means that if $\rho$ commutes with $\rho_A \otimes I_B$, then its quantum discord may be greater than zero. Recently the states having zero quantum discord were discussed in detail \cite{Datta-arxiv, Dakic-arxiv}. However, despite this progress, it is still not known how to compute quantum discord for a general bipartite quantum state due to the complicated maximization procedure. We provide the analytical results to compute the classical correlation and quantum discord for a two-parameter class of states in $2 \otimes d$ quantum systems with $d \geq 3$. This two-parameter class of states is special in the sense that an arbitrary quantum state $\rho$ in $2 \otimes d$ can be reduced to this class provided local operations and classical communication (LOCC) is allowed. This class includes the maximally entangled Bell states, `Werner' states \cite{Wer-PRA89} which include both separable and nonseparable states, as well as others. We have evaluated the analytical expressions for the classical correlation and quantum discord. We have also examined the relation between classical correlation, quantum discord, and entanglement for qubit-qutrit systems.

\section{Quantum discord for two-parameter class of states in a $2 \otimes d$ quantum systems}\label{tp-states}

The class of states with two real parameters $\alpha$ and $\gamma$ in a $2 \otimes d$ quantum system \cite{Chi-JPA36-2003} is given as 
\begin{eqnarray}
\rho_{\alpha, \gamma} =& \alpha \, \sum_{i = 0}^{1} \sum_{j = 2}^{d-1} \, | i\, j \rangle\langle i \, j| + \beta \, ( | \phi^+ \rangle\langle \phi^+| + | \phi^- \rangle\langle \phi^-| + | \psi^+ \rangle\langle \psi^+| \, ) \nonumber \\&  + \gamma \, | \psi^- \rangle\langle \psi^-| \,,
\label{Eq:rhoag}
\end{eqnarray}
where $\{ \, |i \, j \rangle : \, i = 0, \, 1, j = 0, \, 1, \, \ldots \, , \, d-1 \, \}$ is an orthonormal basis for $2 \otimes d$ quantum system and
\begin{eqnarray}
| \, \phi^\pm \rangle &=& \frac{1}{\sqrt{2}} \, ( \, |0\, 0\rangle \pm |	1\,1 \rangle \, ) \\
| \, \psi^\pm \rangle &=& \frac{1}{\sqrt{2}} \, (\, |0\,1 \rangle \pm |1\,0 \rangle ) \, , 
\end{eqnarray}
and the parameter $\beta$ is dependent on $\alpha$ and $\gamma$ by the unit trace condition, 
\begin{eqnarray}
 2 \, ( d - 2) \alpha + 3 \, \beta  + \gamma = 1 \, .
\end{eqnarray}
From Eq.~(\ref{Eq:rhoag}) one can easily obtain the range of parameters as $ 0 \leq \alpha \leq 1/(2 (d-2))$ and $0 \leq \gamma \leq 1$. We note that the states of the form $\rho_{0, \gamma}$ are equivalent to Werner states \cite{Wer-PRA89} in a $2 \otimes 2$ quantum systems. Moreover, the states $\rho_{\alpha,\gamma}$ have the property that their PPT (positive partial transpose) region is always separable \cite{Chi-JPA36-2003}. The \ref{App} describes the procedure of transforming an arbitrary quantum state $\rho$ in $2 \otimes d$ to $\rho_{\alpha, \gamma}$ with the help of local operations and classical communication (LOCC).

Now we calculate the quantum discord for this family of quantum states. It is known that any von Neumann measurement for subsystem $A$ can be written as Ref. \cite{Luo-PRA77-2008}
\begin{eqnarray}
A_i = V \, \Pi_i \, V^\dagger : \quad i = 0,1 \, , \label{Eq:VNmsur}
\end{eqnarray}
where $\Pi_i = |i\rangle\langle i |$ is the projector for subsystem $A$ along the computational base $|i\rangle$ and 
$V \in SU(2)$ is a unitary operator with unit determinant. After the measurement, the state $\rho_{\alpha , \gamma}$ will change to the ensemble $\{ \rho_i , p_i \}$, where
\begin{eqnarray}
\rho_i := \frac{1}{p_i} (A_i \otimes I ) \, \rho_{\alpha , \gamma} \, (A_i \otimes I) \,,
\label{Eq:Tauk}
\end{eqnarray}
and $p_i = \mathrm{tr} \, [\, (A_i \otimes I ) \, \rho_{\alpha , \gamma} \, (A_i \otimes I) \, ]$. The $\{ \rho_i , p_i \}$, with $i=0,1$ are of subsystem $B$ and thus $d \times d$ density matrices.  

We may write any $V \in SU(2)$ as
\begin{eqnarray}
V = t \, I + \mathrm{i} \, \vec{y}\cdot\vec{\sigma} \, ,\label{Eq:defV}
\end{eqnarray}
with $t,y_1,y_2,y_3 \in \mathbb{R}$ and $t^2 + y_1^2 + y_2^2 + y_3^2 = 1$. This implies that these parameters, three among them independent, assume their values in the interval $[-1,1]$, i.\,e. $t\, , \, y_i \in [-1,1]$ for $i = 1,2,3$. 
The ensemble $\{\rho_i , p_i\}$ can be characterized by their eigenvalues \cite{Luo-PRA77-2008,Mali-PRA81-2010}. Interestingly, it turns out that the eigenvalues of the measurement ensemble do not depend on the parameters of $SU(2)$ for any $d$. In a previous study, we have already observed this peculiarity for certain quantum states in $2 \otimes 2$ quantum systems where for example the eigenvalues of the measurement ensemble for Werner states do not depend on $SU(2)$ parameters \cite{Mali-PRA81-2010}. As Werner states are special case of these states, so we have identified a whole class of quantum states in $2 \otimes d$ quantum systems which have this property. The main reason for this behaviour is the highly symmetric nature of these states that is, $U \otimes U$-invariance (See \ref{App}). Therefore, as a result there is no complicated extremization or minimization procedure involved over von Neumann measurements for these states and we can easily obtain the analytical expressions for classical correlation and quantum discord. 

The quantum conditional entropy in Eq.~(\ref{Eq:QCE}) is given as
\begin{eqnarray}
S (\rho_{\alpha , \gamma} |\{A_i\}) = p_0 \, S(\rho_0) + p_1 \, S(\rho_1) \, . \label{Eq:qcetx}
\end{eqnarray}
It turns out that $S(\rho_0) = S(\rho_1)$ and hence, the quantum conditional entropy becomes
\begin{eqnarray}
S (\rho_{\alpha , \gamma} |\{A_i\}) = S(\rho_0) = S(\rho_1) \, . \label{Eq:qcetx2}
\end{eqnarray}

As per Eq.~(\ref{Eq:CC}), the classical correlation is obtained as
\begin{eqnarray}
\mathcal{C}(\rho_{\alpha , \gamma}) &=& \sup_{\{A_i\}} \, [ \, \mathcal{I}(\rho_{\alpha , \gamma}|\{A_i\}) \, ] \nonumber \\ &=& S(\rho_{\alpha , \gamma}^B) - \min_{\{A_i\}} \, [ \, S (\rho_{\alpha , \gamma}|\{A_i\})\, ] \, . \label{Eq:qmimtaux}
\end{eqnarray}
Therefore, to calculate the classical correlation and consequently quantum discord, we have to minimize Eq.~(\ref{Eq:qcetx2}). However, as it is shown below that the eigenvalues of $\rho_i$ do not depend on the parameters of von Neumann measurements, therefore the minimization procedure is not required.

The reduced density matrix $\rho_{\alpha, \gamma}^B$ is $d \times d$ diagonal matrix and hence have $d$ eigenvalues. The spectrum of $\rho_{\alpha, \gamma}^B$ is given as
\begin{eqnarray}
\bigg\{ \frac{3 \, \beta + \gamma }{2} , \, \frac{3 \, \beta + \gamma}{2}, \, 2 \, \alpha , \, 2 \, \alpha , \ldots , \, 2 \, \alpha \bigg\}. \label{Eq:eval}
\end{eqnarray}
The entropy of $\rho_{\alpha, \gamma}^B$ is given as
\begin{eqnarray}
S( \rho_{\alpha, \gamma}^B) = - 2 \, (d-2) \, \alpha \log (2 \, \alpha) - (3 \, \beta + \gamma \, ) \log (\frac{3 \, \beta + \gamma }{2}) \, , \label{Eq:prob}
\end{eqnarray}
where all logrithms in this paper are taken to base $2$. Similarly, the expression for $S (\rho_{\alpha , \gamma} |\{A_i\})$ is given as 
\begin{eqnarray}
S (\rho_{\alpha , \gamma} |\{A_i\}) =& -2 \, (d-2) \, \alpha \, \log(2 \, \alpha) - 2 \, \beta \log (2 \, \beta) \nonumber \\& - (\beta + \gamma ) \, \log(\beta + \gamma) \,.
\label{Eq:thetap}
\end{eqnarray}
The classical correlation is given as
\begin{eqnarray}
\mathcal{C}(\rho_{\alpha,\gamma}) =& - (3 \, \beta + \gamma) \, \log(\frac{3 \, \beta + \gamma }{2}) + 2 \, \beta \, \log(2 \, \beta) \nonumber \\& + (\beta + \gamma) \log(\beta + \gamma) \, . \label{Eq:cc23} 
\end{eqnarray}
The quantum mutual information for $\rho_{\alpha , \, \gamma}$ is given as
\begin{eqnarray}
\mathcal{I} (\rho_{\alpha , \gamma}) &=& S(\rho_{\alpha , \gamma}^A) + S(\rho_{\alpha , \gamma}^B) - S (\rho_{\alpha , \gamma}) \nonumber \\& 
=& (3 \, \beta + \gamma) \, \log(\frac{4}{3 \, \beta + \gamma}) + 3 \, \beta \log(\beta) + \, \gamma \, \log(\gamma) \, , \label{Eq:mi23}
\end{eqnarray}
and quantum discord is simply given as
\begin{eqnarray}
\mathcal{Q}(\rho_{\alpha,\gamma}) &=& \mathcal{I} (\rho_{\alpha,\gamma}) - \mathcal{C}(\rho_{\alpha,\gamma}) \nonumber \\&
=& \beta \, \log(2 \, \beta) + \gamma \, \log(2 \, \gamma) - (\beta + \gamma) \log(\beta + \gamma) \,. 
\label{Eq:qd23}
\end{eqnarray}

We note that the reduced density matrix $\rho_{\alpha,\gamma}^A$ is always equal to maximally mixed state irrespective of the dimension of second Hilbert space, that is, $\rho_{\alpha,\gamma}^A = I/2$, whereas $\rho_{\alpha,\gamma}^B$ is equal to maximally mixed state only for $d = 2$, that is, for qubit-qubit system. For $d \geq 3$, qubit marginal is maximally mixed and qudit marginal is not. Thus, we have obtained analytically the classical correlation and thereby quantum discord for two-parameter class of states in qubit-qudit systems. For the special case of two-qubit X-states \cite{Rau-JPA09} with restrictions $\rho_{11} = \rho_{44}$, $\rho_{22} = \rho_{33}$, and with real off-diagonal elements, we recover the results of Refs. \cite{Luo-PRA77-2008,Mali-PRA81-2010}. Hence our study is generalization of some of the results obtained earlier and the extension of all previous studies on quantum discord to qubit-qudit quantum systems.

\section{Discord and entanglement for qubit-qutrit system}\label{relation}

In this section, we study the relation between the classical correlation, quantum discord, and entanglement for initial qubit-qutrit states. The two-parameter states for this dimension of Hilbert space are give as
\begin{eqnarray}
\sigma_{\alpha, \gamma} =& \alpha \, (\, | 0\, 2 \rangle\langle 0 \, 2| + |1 \, 2 \rangle\langle 1 \, 2 | \, ) + \beta \, ( \, | \phi^+ \rangle\langle \phi^+| + \nonumber \\& | \phi^- \rangle\langle \phi^-| + | \psi^+ \rangle\langle \psi^+| \, ) + \gamma \, | \psi^- \rangle\langle \psi^-| \,.
\label{Eq:rhoag23}
\end{eqnarray}

The negativity \cite{Vidal-PRA65-2002} is a measure of entanglement which completely characterizes and quantifies the set of entangled states for $2 \otimes 2$ and $2 \otimes 3$ systems \cite{Alber-QI2001,Horodecki-RMP-2009}. However, the negativity completely characterizes and quantifies the entanglement of this family of quantum states in all dimensions of Hilbert space because it was shown \cite{Chi-JPA36-2003} that the PPT region of these states is always separable and there is no possibility of bound entangled states \cite{Horodecki-RMP-2009} in this family. The negativity for $\rho_{\alpha,\gamma}$ is given as \cite{Chi-JPA36-2003} 
\begin{eqnarray}
N(\rho_{\alpha,\gamma}) = \max \, \{ \, 0, \, 2 \, (d-2) \, \alpha + 2 \, \gamma -1 \, \} \, . 
\end{eqnarray}
For qubit-qutrit system negativity would be $N(\sigma_{\alpha,\gamma}) = \max \, \{ \, 0, \, 2 \, \alpha + 2 \, \gamma -1 \, \}$. 

Let us consider few initial states as examples to study classical correlation, entanglement and quantum discord.

$(1):$ As a trivial example we consider the initial state Eq.(\ref{Eq:rhoag23}) with $\alpha = \beta = 0$, and $\gamma = 1$. Eqs.(\ref{Eq:cc23}), (\ref{Eq:mi23}), and (\ref{Eq:qd23}) predict that
\begin{eqnarray}
\mathcal{I}(\rho_{0,1}) = 2 \, , \, \, \, \mathcal{C}(\rho_{0,1}) = \mathcal{Q}(\rho_{0,1}) = N(\rho_{0,1}) = 1 \, ,  
\end{eqnarray}
which is obviously the case for maximally entangled Bell state \cite{Groisman-PRA72-2005}, where entanglement and quantum discord also coincide and are identical.

$(2):$ As a second example consider the initial states Eq.\,(\ref{Eq:rhoag23}) with $\alpha = \gamma = 0$ and $\beta = 1/3$. Clearly the negativity of $\rho_{0,0}$ is zero, that is $N(\rho_{0,0}) = 0$ and the states have positive partial transpose (PPT) and hence separable. Although there is no entanglement in these states for this choice of two parameters, nevertheless, the states were constructed by a fraction of maximally entangled state $|\psi^+\rangle$ and we expect some nonclassical correlation in it. The classical correlation and quantum discord are given as
\begin{eqnarray}
\mathcal{C}(\rho_{0,0}) = \frac{5}{3} + \log(\frac{1}{3}) \, , \, \, \, \mathcal{Q}(\rho_{0,0}) = \beta = \frac{1}{3} \, , 
\end{eqnarray}
respectively. Hence, both classical correlation and quantum discord are strictly positive.

$(3):$ Consider the initial states Eq.\,(\ref{Eq:rhoag23}) with $\gamma = 0$. Again the negativity for $\rho_{\alpha , 0}$ is zero, that is, $N(\rho_{\alpha,0}) = 0$ and hence the states are PPT and separable. Nevertheless, the classical correlation is given as
\begin{eqnarray}
\mathcal{C}(\rho_{\alpha,0}) =& -(1 - 2\,\alpha) \log(\frac{1 - 2 \, \alpha}{2}) + \frac{2(1 - 2 \, \alpha)}{3} \nonumber \\& \times \, \log(\frac{2(1 - 2 \, \alpha)}{3}) + \frac{1 - 2 \, \alpha}{3} \log(\frac{1 - 2 \, \alpha}{3}) \, , 
\end{eqnarray}
and quantum discord is given as
\begin{eqnarray}
\mathcal{Q}(\rho_{\alpha,0}) = \frac{1 - 2 \, \alpha}{3} \,. 
\end{eqnarray}

\begin{figure}[h]
\scalebox{2.0}{\includegraphics[width=1.7in]{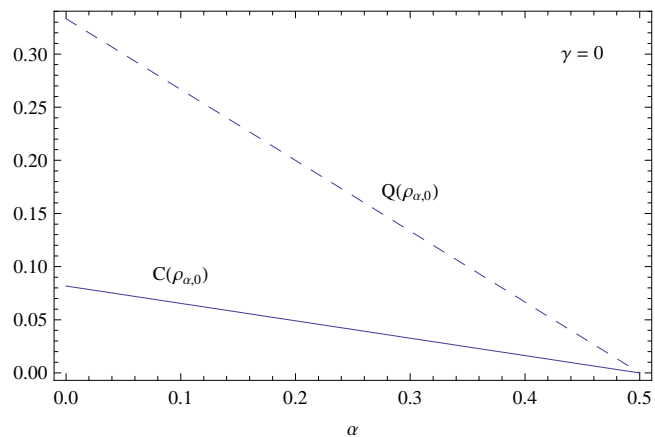}}
\centering
\caption{Quantum discord and classical correlation for the class of states in Eq.(\ref{Eq:rhoag23}) with $\gamma = 0$ and $d = 3$, i.e., qubit-qutrit quantum system of the adjoining text are plotted for $\rho_{\alpha,0}$.}
\label{Fig:1}
\end{figure}

Figure \ref{Fig:1} displays the classical correlation and quantum discord for $\rho_{\alpha , 0}$ against the parameter $\alpha$. The solid line presents classical correlation, whereas the dashed line is for quantum discord. It can be seen that for this particular initial state, quantum discord is always greater than classical correlation.

$(4):$ Let us consider the initial state with parameter $\beta = 0$. It can be seen easily that initial states consist of maximally entangled state mixed with noisy component, that is, for $\gamma = 1$, the state is maximally entangled and it is separable only for $\gamma = 0$. It turns out that the negativity, classical correlation, and quantum discord are equal for this state and given as
\begin{eqnarray}
N(\rho_{\alpha,\gamma}|_{\beta = 0}) &=& \mathcal{C}(\rho_{\alpha,\gamma}|_{\beta = 0}) = \mathcal{Q}(\rho_{\alpha,\gamma}|_{\beta = 0}) \nonumber \\&
=& 1 - 2 \, \alpha \, . 
\end{eqnarray}
As $\alpha \in [0, \, 1/2]$, hence for $\alpha = 0$, these correlations achieve their maximum value equal to $1$ and they are zero for $\alpha = 1/2$ which corresponds to value of parameter $\gamma = 0$.

\begin{figure}[h]
\scalebox{2.0}{\includegraphics[width=1.7in]{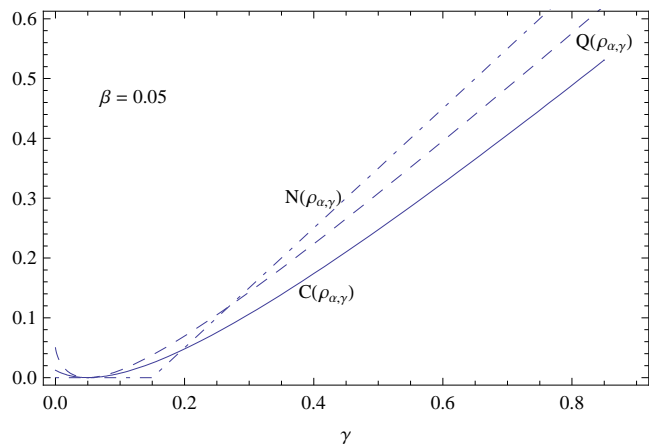}}
\centering
\caption{As in Fig. $1$ for the class of states in Eq.(\ref{Eq:rhoag23}) for qubit-qutrit quantum systems in the text. Top curve (dotted-dashed line) is for negativity, middle one (dashed) is for quantum discord, and bottom curve (solid line) is for classical correlation.}
\label{Fig:2}
\end{figure}

$(5):$ Finally, we consider the initial states with $\alpha, \, \beta , \, \gamma > 0$. The classical correlation, quantum discord, and negativity have been plotted in Figure \ref{Fig:2} for $\rho_{\alpha , \gamma}$ for a particular value of the parameter $\beta = 0.05$. The solid line is for classical correlation, the dotted-dashed line is for negativity, and the dashed line for quantum discord. For this particular initial state, the classical correlation is always smaller than quantum discord, however it is larger than entanglement only for a particular range and then becomes always smaller than negativity. Similarly, quantum discord is also larger than negativity for some range of parameter $\gamma$ and then becomes smaller.

\begin{figure}[h]
\scalebox{2.0}{\includegraphics[width=1.7in]{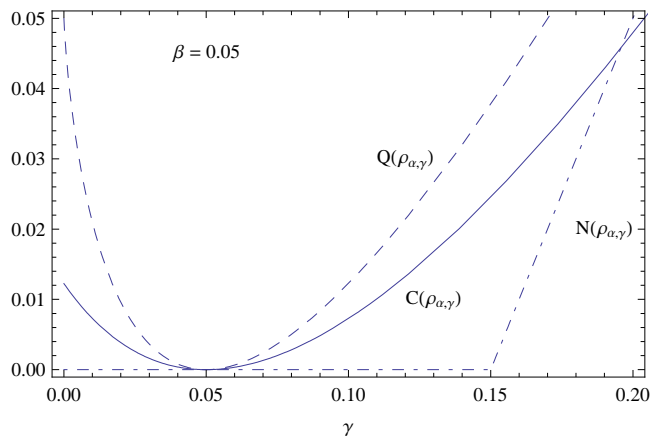}}
\centering
\caption{Enlarged plot for given range of Fig. $2$ for the class of states in Eq.(\ref{Eq:rhoag23}) for qubit-qutrit quantum systems. The dotted-dashed line is for negativity, dashed line is for quantum discord, and solid line is for classical correlation. The figure shows that for $\beta = \gamma$, all correlations are equal to zero.}
\label{Fig:3}
\end{figure}

We note that both classical correlation and quantum discord vanishes for a particular value of $\gamma$ as shown in Figure \ref{Fig:2}. We show the plot for this range of values in Figure \ref{Fig:3}. It can be seen that for $\beta = \gamma$, classical correlation, quantum discord, and entanglement vanishes. It might appear as surprising, however it is obvious that the initial states with $\beta = \gamma$ are completely uncorrelated states and therefore we expect that all correlation measures must be zero.


\section{Summary}\label{conclusion}

We have derived analytical expressions for the classical correlation and quantum discord for two-parameter class of states in $2 \otimes d$ quantum systems. This class has a peculiarity that the eigenvalues of its measurement ensemble do not depend on the parameters of von Neumann measurements. These highly symmetric quantum states enable us to calculate the corresponding expressions for classical correlation and quantum discord without any complicated maximization/minimization procedure. Another important aspect is the fact an arbitrary state in $2 \otimes d$ quantum system can be transformed into this class of two-parameters states by LOCC. Hence, if LOCC are allowed then one can find the classical correlation and quantum discord for qubit-qudit quantum systems for any arbitrary quantum state. These results also generalizes some of the results previously available only for a single and three-parameter subsets of such states in qubit-qubit Hilbert space. Various correlations such as the classical correlation, quantum discord, and entanglement can now be examined for a larger class of bipartite quantum states that includes maximally or partially entangled states, and mixed states that are separable or non-separable. Perhaps the investigations on quantum discord or other quantum correlation may reveal the true nature of nonlocality and nonclassicality in a more clear way. We conclude that entanglement and quantum discord are not identical correlation and quantum discord is a fundamentally different resource than entanglement. More rigorous studies on unified view of classical and quantum correlation is desired.

During the completion of this work, I have learned that similarly studies are in progress \cite{SV-Rau-prog} on evaluating quantum discord for qubit-qudit quantum systems in a totally different way and for more general class of quantum states.

\ack
I thank Prof. A. R. P. Rau for reading the manuscript and giving useful suggestions. I am also thankful to Prof. Gernot Alber for his comments and his kind hospitality at Technische Universit\"at Darmstadt where the final part of this work is done.

\appendix

\section{Transformation of an arbitrary state into two-parameter class by LOCC}\label{App}

Now we briefly describe the process of transforming an arbitrary state in $2 \otimes d$ quantum system to a state of the form $\rho_{\alpha , \gamma}$. All the subsequent discussion has been described in Ref. \cite{Chi-JPA36-2003}. We reproduce the arguments here for the convenience of readers. It was shown that an arbitrary state in a $2 \otimes d$ quantum system can be transformed to a state of the form $\rho_{\alpha, \gamma}$ (Eq.(\ref{Eq:rhoag})) by LOCC. The states $\rho_{\alpha, \gamma}$ are invariant under all unitary operations of the form $U \otimes U$ on a $2 \otimes d$ quantum system. Let $U(k)$ be the group of all unitary operators on a $k$-dimentional Hilbert space, and $\{ \, |0\rangle_A \, |1\rangle_A \, \}$ and $\{ \, |0\rangle_B \, |1\rangle_B \,\ldots \, , |d-1\rangle_B \, \}$ be bases of $\mathcal{H}_A$ and $\mathcal{H}_B$, respectively. For ease, we identify a unitary operator $U_A \in U(2)$ with $U_B \in U(d)$ if for $j = 0, \, 1,$ $U_A |j\rangle_A = a_j |0\rangle_A + b_j |1\rangle_A$ and $U_B |j\rangle_B = a_j |0\rangle_B + b_j |1\rangle_B$. For $0 < m < d$, we let
\begin{eqnarray}
 G(m,d) = \{ \, U \in U(d): U (\mathcal{H}_m) = \mathcal{H}_m , \, U(\mathcal{H}_m^\perp) = \mathcal{H}_m^\perp \, \}
\end{eqnarray}
where $\mathcal{H}_m$ is a subspace of $\mathcal{H}_B$ generated by $|0\rangle_B , \, |1\rangle_B , \, \ldots \, |m-1\rangle_B$, and $\mathcal{H}_m^\perp$ is the orthogonal complement of $\mathcal{H}_m$ in $\mathcal{H}_B$. Then $G(2,d)$ is a subgroup of $U(d)$, and if $U$ is a unitary operator in $G(2,d)$ then it is compatible to write a unitary operator of the form $U \otimes U$ on a $2 \otimes d$ quantum system.

The technique of transforming an arbitrary state $\rho$ to $\rho_{\alpha , \gamma}$ by using local operations and classical communication (LOCC) is similar to that presented by Bennet {\it et \, al} \cite{Bennet-PRA54-1996} and D\"ur {\it et al} \cite{Duer-PRA61-2000}. We outline here the main arguement from Ref. \cite{Chi-JPA36-2003}. It was shown that there exist unitary operators $U_k$ and probabilities $p_k$ such that 
\begin{eqnarray}
\sum_k \, p_k \, (\, U_k \otimes U_k \, ) \rho \, ( \, U_k^\dagger \otimes U_k^\dagger \, ) = \rho_{\alpha, \gamma} \,.
\end{eqnarray}

Define the operation $U_\theta$ as $U_\theta : \, | j \rangle \mapsto (\mathrm{e}^{\mathrm{i}\theta})^j \, | j \rangle$, where $\mathrm{i} = \sqrt{-1}$. First perform $U_\pi \otimes U_\pi$ with probability $1/2$ and no operation (identity operation) with probability $1/2$, that is,
\begin{eqnarray}
\frac{1}{2} \, ( \, U_\pi \otimes U_\pi \, ) \rho ( \, U_\pi^\dagger \otimes U_\pi^\dagger \, ) + \frac{1}{2} \, \rho \, . \label{Eq:op1}
\end{eqnarray}
Now define the operation $U_k$ by $U_k : |j\rangle \mapsto (-1)^{\delta_{j,k}} |j\rangle$ for $k = 2, \, 3, \ldots \,, d-1 $, and then for each $k$, perform $U_k \otimes U_k$ with probability $1/2$, while no operation with probability $1/2$, respectively. Here, $U_k \otimes U_k = I \otimes U_k$. Now perform $U_{\pi/2} \otimes U_{\pi/2}$ with probability $1/2$ as in Eq.(\ref{Eq:op1}) and then perform the swap operator $U_{01}: |0\rangle \leftrightarrow |1\rangle (\, |j\rangle \mapsto |j\rangle$ for $2 \leq j \leq d-1\,) $ with probability $1/2$. After these operations, a state of the following form is obtained
\begin{eqnarray}
\sum_{j=2}^{d-1} \, a_j (\,|0 \, j\rangle\langle 0,\,j| + |1\,j \rangle\langle 1\,j| \, ) + b (\, |\phi^+\rangle\langle \phi^+| + \nonumber \\ |\phi^-\rangle\langle \phi^-|\, ) + c_+ | \psi^+ \rangle\langle \psi^+ | + c_- | \psi^- \rangle\langle \psi^-| \,.
\label{Eq:op2} 
\end{eqnarray}
Let $T$ be the unitary operator defined as $|0\rangle \mapsto |0\rangle$, $|1\rangle \mapsto |1\rangle$, $|2\rangle \mapsto |3\rangle$, $|3\rangle \mapsto |4\rangle$, $\ldots$ , $|d-2\rangle \mapsto |d-1\rangle$ and $|d-1\rangle \mapsto |2\rangle$. Now perform the operation:
\begin{eqnarray}
\rho \mapsto \frac{1}{d-2} \sum_{j=0}^{d-3} ( \, T^j \otimes T^j \, ) \rho (\,T^j \otimes T^j \, )^\dagger \, .
\label{Eq:op3}
\end{eqnarray}
Here, $T^j \otimes T^j = I \otimes T^j$ for any $j = 0, \, 1, \, \ldots \, , d-3$. As a result a state in Eq.(\ref{Eq:op2}) becomes
\begin{eqnarray}
a \, \sum_{i = 0}^1 \, \sum_{j=2}^{d-1} \,  |i \, j\rangle\langle i,\,j| + b (\, |\phi^+\rangle\langle \phi^+| + |\phi^-\rangle\langle \phi^-|\, ) \nonumber \\ + c_+ | \psi^+ \rangle\langle \psi^+ | + c_- | \psi^- \rangle\langle \psi^-| \,. \label{Eq:op4} 
\end{eqnarray}
Let $H$ be the unitary operator (Hadamard operator) defined as $|0\rangle \mapsto (|0\rangle + |1\rangle)/\sqrt{2}$
, $|1\rangle \mapsto (|0\rangle - |1\rangle)/\sqrt{2}$ and $|j\rangle \mapsto |j\rangle$ for $2 \leq j \leq d-1$. After performing the following operation as:
\begin{eqnarray}
\rho \mapsto \frac{2}{3} ( \, H \otimes H \, ) \rho (\, H \otimes H \,) + \frac{1}{3} \rho \, ,
\end{eqnarray}
perform the sequence of the previous operations again. One can easily check that Eq.(\ref{Eq:rhoag}) is obtained with two parameters with $\alpha = \sum_{i,j} a_{ij}/(2\,d-4)$ and $\gamma = c_-$. One can also show \cite{Chi-JPA36-2003} that $\rho_{\alpha , \gamma}$ is invariant under all $U \otimes U$, that is, for any $U \in G(2,d)$,
\begin{eqnarray}
 (\, U \otimes U \, ) \rho_{\alpha , \gamma} (\, U^\dagger \otimes U^\dagger \, ) = \rho_{\alpha , \gamma} \, .
\end{eqnarray}


\section*{References} 

\begin{thebibliography}{99}

\bibitem{Alber-QI2001} Alber G, Beth T, Horodecki M, Horodecki P, Horodecki R, R\"otteler M, Weinfurter H, Werner R and Zeilinger A 2001 {\it Quantum Information} (Berlin: Springer-Verlag) ch. 5

\bibitem{Horodecki-RMP-2009} Horodecki R, Horodecki P, Horodecki M and Horodecki K 2009 \RMP {\bf 81} 865

\bibitem{NC-QIQC-2000} Nielsen M A and Chuang I L 2000 {\it Quantum Computation and Quantum Information} (Cambridge: Cambridge Univ. Press)

\bibitem{Bennett-PRA59-1999} Bennett C H, DiVincenzo D P, Fuchs C A, Mor T, Rains E, Shor P W, Smolin J A and Wootters W K 1999 \PR A {\bf 59} 1070

\bibitem{Horodecki-PRA71-2005} Horodecki M, Horodecki P, Horodecki R, Oppenheim J, Sen A, Sen U and Synak-Radtke B 2005 \PR A {\bf 71} 062307

\bibitem{Niset-PRA74-2006} Niset J and Cerf N J 2006 \PR A {\bf 74} 052103

\bibitem{Piani-PRL100-2008} Piani M, Horodecki P and Horodecki R 2008 \PRL {\bf 100} 090502

\bibitem{Piani-PRL102-2009} Piani M, Christandl M, Mora C E and Horodecki P 2009 \PRL {\bf 102} 250503 

\bibitem{Braunstein-PRL83-1999} Braunstein S L, Caves C M, Jozsa R, Linden N, Popescu S and Schack R 1999 \PRL {\bf 83} 1054

\bibitem{Meyer-PRL85-2000} Meyer D A 2000 \PRL {\bf 85} 2014

\bibitem{Datta-PRL100-2008} Datta A, Flammia S T and Caves C M 2005 \PR A {\bf 72} 042316; Datta A and Vidal G 2007 \PR A {\bf 75} 042310; Datta A, Shaji A and Caves C M 2008 \PRL {\bf 100} 050502

\bibitem{Lanyon-PRL101-2008} Lanyon B P, Barbieri M, Almeida M P and White A G 2008 \PRL {\bf 101} 200501

\bibitem{Cui-JPA43-2010} Cui J and Fan H 2010 \JPA {\bf 43} 045305

\bibitem{Modi-PRL104-2010} Modi K, Paterek T, Son W, Vedral V and Williamson M 2010 \PRL {\bf 104} 080501

\bibitem{Groisman-PRA72-2005} Groisman B, Popescu S and Winter A 2005 \PR A {\bf 72} 032317

\bibitem{Schumacher-PRA74-2006} Schumacher B and Westmoreland M D 2006 \PR A {\bf 74} 042305

\bibitem{Ollivier-PRL88-2001} Ollivier H and Zurek W H 2001 \PRL {\bf 88} 017901

\bibitem{Vedral-et-al} Henderson L and Vedral V 2001 \JPA {\bf 34} 6899; Vedral V 2003 \PRL {\bf 90} 050401; Maziero J, Cel\'eri L C, Serra R M and Vedral V 2009 \PR A {\bf 80} 044102

\bibitem{Luo-PRA77-2008} Luo S 2008 \PR A {\bf 77} 042303

\bibitem{Mali-PRA81-2010} Ali M, Rau A R P and Alber G 2010 \PR A {\bf 81} 042105

\bibitem{Li-Luo-PRA-2008} Li N and Luo S 2007 \PR A {\bf 76} 032327; Luo S 2008 \PR A {\bf 77} 022301

\bibitem{Oppenheim-PRL89-2002} Oppenheim J, Horodecki M, Horodecki P and Horodecki R 2002 \PRL {\bf 89} 180402

\bibitem{Kaszlikowski-PRL101-2008} Kaszlikowski D, Sen A, Sen U, Vedral V and Winter A 2008 \PRL {\bf 101} 070502

\bibitem{Dillenschneider-PRB78-2008} Dillenschneider R 2008 \PR B {\bf 78} 224413

\bibitem{Sarandy-PRA80-2009} Sarandy M S 2009 \PR A {\bf 80} 022108

\bibitem{Werlang-PRA80-2009} Werlang T, Souza S, Fanchini F F and Villas Boas C J 2009 \PR A {\bf 80} 024103

\bibitem{Fanchini-PRA81-2010} Fanchini F F, Werlang T, Brasil C A, Arruda L G E and Caldeira A O 2010 \PR A {\bf 81} 052107; Maziero J, Werlang T, Fanchini F F, Cel\'eri L C and Serra R M 2010 \PR A {\bf 81} 022116

\bibitem{Bylicka-PRA81-2010} Bylicka B and Chru\'sci\'nski D 2010 \PR A {\bf 81} 062102

\bibitem{Ferraro-PRA81-2010} Ferraro A, Aolita L, Cavalcanti D, Cucchietti F M and Acin A 2010 \PR A {\bf 81} 052318

\bibitem{Datta-arxiv} Datta A 2010 ({\it Preprint} arXiv: quant-ph/1003.5256)

\bibitem{Dakic-arxiv} Dakic B, Vedral V and Brukner C 2010 ({\it Preprint} arXiv:quant-ph/1004.0190)

\bibitem{Wer-PRA89} Werner R F 1989 \PR A {\bf 40} 4277

\bibitem{Chi-JPA36-2003} Chi D P and Lee S 2003 \JPA {\bf 36} 11503

\bibitem{Rau-JPA09} Rau A R P 2009 \JPA {\bf 42} 412002

\bibitem{Vidal-PRA65-2002} Vidal G and Werner R F 2002 \PR A {\bf 65} 032314

\bibitem{Bennet-PRA54-1996} Bennet C H, DiVincenzo D P, Smolin J A and Wootters W K 1996 \PR A {\bf 54} 3824

\bibitem{Duer-PRA61-2000} D\"ur W, Cirac J I, Lewenstein M and Bru\ss \, D 2000 \PR A {\bf 61} 062313

\bibitem{SV-Rau-prog} Vinjanampathy S and Rau A R P 2010 ({\it in progress})

\end{thebibliography}
\end{document}